\documentclass[twocolumn,aps,superscriptaddress,pra]{revtex4}
\usepackage{amsmath}%
\usepackage{amsfonts}%
\usepackage{amssymb}%
\usepackage{graphicx}
\begin{document}
\bibstyle{plain}

\title{$N$-particle nonclassicality without $N$-particle correlations}
\author{Marcin Wie\'sniak}
\author{Mohamed Nawareg}
\author{Marek \.Zukowski}
\affiliation{Institute of Theoretical Physics and Astrophysics, University of Gda\'nsk, PL-80-952 Gda\'nsk,Poland}
\begin{abstract} 
Most of known multipartite Bell inequalities involve correlation functions for all subsystems. They are useless for entangled states without such correlations. We give a method of derivation of families of  Bell inequalities for $N$ parties, which involve, e.g., only $(N-1)$-partite correlations, but still are able to detect proper N-partite entanglement. We present an inequality  which reveals five-partite entanglement despite only four-partite correlations. Classes of inequalities introduced here can be put into a handy form of a single non-linear inequality.  An example  is given of an $N$ qubit state, which strongly violates such an inequality, despite having no $N$-qubit correlations. This surprising property might be of potential value for quantum information tasks.
\end{abstract}

\maketitle
\section{Introduction}
Entanglement might be seen as the essential feature of quantum mechanics \cite{SCHROEDINGER}. 
Using maximally entangled states Einstein, Podolsky, and Rosen, \cite{epr}, tried to argue that quantum theory should be supplemented with `elements of reality' respecting locality. This was disproved by  Bell \cite{bell}, who derived an inequality, which is satisfied by all local realistic theories, and showed that it is not satisfied for a quantum process involving entanglement.

As the studies of entanglement, and its applications progressed, Bell inequalities were further generalized to more complicated experimental situations (see, e.g., \cite{PAN} for multiphoton interference effects), and found direct applications in quantum cryptography, reduction of communication complexity, and other quantum information protocols. They also serve as device independent entanglement indicators (witnesses). 

After the discovery of the Greenberger-Horne-Zeilinger (GHZ) correlations \cite{GHZ}, series of new inequalities for an arbitrarily number of qubits, with two alternative observables per side were derived in Refs. \cite{mermin, ardehali,kb,ww,wz,zb}. Basing on similar ideas inequalities with more settings per site were derived in Refs. \cite{chen,lpzb}. These Bell inequalities are for correlation functions for all subsystems. 

However, it has been recently shown that for an odd $N$ one can find $N$-qubit states, which, despite being $N$-partite entangled, do not posses any correlations between all  $N$ qubits \cite{dag}. Still, they have non-zero correlations between all subsets of $N-1$  parties. Such states are now within the experimental reach, for Dicke states see Ref. \cite{dicke1} or for rotationally invariant states see Ref. \cite{singlets}. Naturally, as there are no correlations between all qubits, such states cannot violate any of the aforementioned inequalities, as they involve correlations of the highest order. Additionally, even if they violate $(N-1)$-partite families mentioned above, this would indicate only $(N-1)$-partite entanglement. Therefore one is tempted to think that such states are useless for a direct application in quantum information tasks involving {\em all} observers.  

Here, we show an algorithm to derive Bell inequalities involving only  correlation functions of lower order, for all $N-1$ partners subsets, however, altogether involving all $N$ parties. The inequalities presented here use lower order correlations in a symmetric way. That is, e. g., for five observers, correlations involving each of five subsets of four of them are used in such inequalities. Thus in effect we have all five partners on equal footing. Violation of such inequality could be called five-particle non-classicality without five-particle correlations.
They form a family totally unrelated to earlier inequalities involving just one specific subset of observers. We find a state possessing no $N=5$ partite correlations, which violates  an inequality of the type we present in this paper. This property reveals some kind of hidden non-classical five-particle correlations: they involve subsets of four partners, but the set of the subsets contains all of them. 

Thus such states, which at first glance seem to be worthless in for five partite quantum protocols,  definitely could be useful in e.g. some communication complexity reducing protocols. This is justified by the findings of Ref.  \cite{BRUKNER} in which was shown that with any Bell inequality violated by a quantum state (customarily called a relevant inequality), one can link a communication complexity protocol for which employment of quantum methods gives an advantage over all possible classical protocols. 

Our approach is presented for a multiparty scenario in which every observer is allowed to perform one of two dichotomic measurements (that is two-valued ones). Thus we use qubits as natural quantum systems for which we seek violations of the new inequalities, but of course the systems could be more complicated. 

\subsection{Bell inequalities}
We present a short introduction to the notion of Bell inequalities. Let us start with the case of two observers. Consider pairs of photons simultaneously emitted in opposite directions. They
arrive at two very distant measuring devices A and B, operated by Alice and Bob, respectively.
Their apparatuses have a "knob", which specifies
which dichotomic (i.e., two-valued, yes-no) observable they
measure. {E.g., for a device consisting of a polarizing beamsplitter
and two detectors behind its outputs, this knob specifies the orientation of
the polarizer, etc.} One can assign to the two possible results (eigen-)
values $\pm1$ (for yes or no). {We assume that we have a perfect
situation in which the detectors never fail to register a photon.} Alice and
Bob are at any time  \emph{free} to choose the
knob settings (also in a \textquotedblleft delayed
choice\textquotedblright\ mode, after an emission). In other words,  the process governing such choices is stochastically independent of the other ones in the experiment.

Any classical description of such a scenario, consistent with relativity, would be equivalent to {\em local realism}, or expressible by a local realistic model. 
Assuming {\em realism} as the theoretical description allows one to introduce values of results of all possible experiments that
can be performed on an individual system, without caring which measurement is actually done on the system. Note that this assumption is in conflict with complementarity (``unperformed experiments have no results'',  Peres \cite{PERES}). According to \textit{locality},
the random choices and the observations made by Alice and Bob, which can be simultaneous in certain reference
frames, cannot influence each other, and the choice made on one side cannot
influence the results on the other side, and \emph{vice versa}. 

To test local realism Alice chooses randomly, with equal probability,
to measure either observable $\hat{A}_{1}^1$ or $\hat{A}_{2}^1$, and Bob either
$\hat{A}_{1}^2$ or $\hat{A}_{2}^2$.  Let us denote the hypothetical results that
they may get for an emitted pair by $A_{1}^{1}$ and $A_{2}^{1}$ for Alice's
two possible choices, and $A_{1}^{2}$ and $A_{2}^{2}$ for Bob's. {Note that if the results were depending also on the settings on the other side, one would  use a two-index notation $A_{a,b}^{j}$, with $a,b=1,2$, where $a $ and $b$ indicate Alice's and Bob's  choices of the settings (who are indicated by $j=1,2$). As the index of the distant measuring station is missing, this is tantamount to the locality assumption.}  The numerical
values of these can be $\pm1$. The assumption of local realism allows one to
treat $A_{1}^{j}$ and $A_{2}^{j}$ on equal footing as two numbers, one of them
revealed in the experiment, the second one unknown, but still $\pm1$. Thus,
their sum and difference always exist, and is an algebraic expression. For these values for the possible  measurement results one has either $|A_{1}^{j}-A_{2}^{j}|=2$
and $|A_{1}^{j}+A_{2}^{j}|=0$, or $|A_{1}^{j}-A_{2}^{j}|=0$ and $|A_{1}
^{j}+A_{2}^{j}|=2$. Thus, the following trivial
relation holds
\begin{equation}
\sum_{s_{1}=\pm1}\sum_{s_{2}=\pm 1}S(s_{1},s_{2})(A_{1}^{1}+s_{1}A_{2}
^{1})(A_{1}^{2}+s_{2}A_{2}^{2})=\pm 4, \label{trivialrelation}
\end{equation}
where $S(s_{1},s_{2})$ is an arbitrary \textquotedblleft
sign\textquotedblright\ function of $s_{1}$ and $s_{2}$, that is always
$|S(s_{1},s_{2})|=1.$ 

Imagine now that $K$ pairs of photons are emitted, pair
by pair, and {$K$ is sufficiently large, $\sqrt{1/K}\ll 1$}. The average value
of the products of the local values for a joint test (the Bell correlation
function), during which, for all photon pairs, only {\em one} specific pair of observables
(say, $\hat{A}_n^1$ and $\hat{A}_{m}^{2}$) is chosen, is given by

\begin{equation}
\langle A_{n}^1A_{m}^2\rangle=\frac{1}{K}\sum_{j=1}^{j=K}A_{n}^{1}A_{m}^{2},\label{BELLEQ-0}%
\end{equation}
where $n=1,2$ and $m=1,2$ (we have dropped the additional index $j$ denoting the $j$th pair). The relation (\ref{BELLEQ-0}) after averaging, and using the assumption that the choice of measurement settings is independent of all other features of the experiment (which allows one to claim that the averages for the part of the ensemble in which specific settings were actually chosen, cannot differ much from the averages for the theoretical prediction for the full ensemble (\ref{BELLEQ-0})), 
implies that for the four \textit{possible} choices, if local realism holds,  the following
inequality must be satisfied (all such inequalities can be reduced to just
a single non-linear one)(see \cite{wz}, \cite{ww}):
\begin{align}
&  \sum_{s_{1}=\pm1}\sum_{s_{2}=\pm1}\big|\langle A_{1}^1A_{1}^2+s_{2}A_{1}^1A_{2}^2\nonumber\\
&  +s_{1}A_{2}^1A_{1}^2+s_{1}s_{2}A_{2}^1A_{2}^2\rangle\big|\leq4.\label{BELLEQ-1}%
\end{align}
If in  (\ref{trivialrelation}) 
one chooses a non-factorisable function $S(s_{1},s_{2})$, say, $\frac{1}{2}%
(1+s_{1}+s_{2}-s_{1}s_{2})$, after averaging the famous Clauser-Horne-Shimony-Holt (CHSH) inequality \cite{CHSH} is recovered (the simple algebra to reach this
result rests on the fact that $\sum_{{s_{j}}=\pm1}s_{j}=0$, while
$\sum_{{s_{j}}=\pm1}s_{j}^{2}=2$)
\begin{align}
S_{\mathrm{Bell}} &  \equiv\left\vert\langle A_{1}^1A_{1}^2+A_{1}^1A_{2}^2\rangle\right.
\nonumber\\
&  \left.  \left.  +\langle A_{2}^1A_{1}^2-A_{2}^1A_{2}^2\rangle\right\vert \leq2.\right.
\label{BELLINEQ}%
\end{align}

Some quantum processes involving entangled states violate this inequality. For example the
predictions for the spin-$1/2$ singlet give correlations for which
$S_{\mathrm{Bell}}$ can acquire the maximal value $2\sqrt{2}$. This is the root of applicability of quantum correlations in solutions of many classically impossible problems, like secure cryptographic key distribution, 
reduction of communication complexity, truly random number generators, etc. Note that violation of Bell inequalities, could be translated into the following computational situation. We have two partners, each possessing a supercomputer, they can share whatever programs they want, exchanging as much information as possible. However, at certain moment the information exchange is cut. Only then, each of them tosses a coin. Heads mean $1$, tails $2$. Alice's result is fed to her computer, Bob's into his one. Basing on these inputs, the computers are to produce (without any inter-computer communication), by whatever programming code, a number $+1$, or $-1$. The procedure is repeated $K$ times. Such a scheme, no matter how expensive and complicated, will never reproduce average products of local results predicted for some entangled two qubit states, and a pair of spatially separated observers, choosing between certain two dichotomic local measurements on their qubits (as in the thought experiment above). Thus, either computer simulations of a pair of qubits require additional communication, which is never cut, or the very idea that a local quantum system can be simulable via a local computer is wrong.


\emph{Multiparticle Bell inequalities.  } The systematic research on multiparty Bell inequalities started after the discovery of GHZ correlations. Series of such
inequalities were discovered by Mermin \cite{mermin}, Ardehali \cite{ardehali},and Belinskii
and Klyshko  \cite{kb}. To get the full set of such inequalities, for $N$ partite correlation
functions involving the product of the result of {\em all} parties, it is enough to
generalize the relation (\ref{BELLEQ-0}) to the situation in question. E.g.,
extending (\ref{trivialrelation}) for three partners one has
\begin{eqnarray}
&  \sum_{s_{1},s_{2},s_{3}=\pm 1}S(s_{1},s_{2},s_{3})(A_{1}^{1}+s_{1}A_{2}^{1})
(A_{1}^{2}+s_{2}A_{2}^{2})&\nonumber\\
&  \times(A_{1}^{3}+s_{3}A_{2}^{3})=\pm8.& \\ \nonumber
\end{eqnarray}
This leads to the following Bell inequality \cite{ww}, \cite{wz}:
\begin{equation}
\sum_{s_{1},s_{2},s_{3}=\pm1}|\sum_{k,l,m=1,2}s_{1}s_{2}s_{3}\langle A_{k}^1
A_{l}^2A_{m}^3\rangle|\leq8,\label{BELLEQ-4}
\end{equation}
which is the \emph{necessary and sufficient} condition for correlation
functions involved in it, $\langle A_{k}^1A_{l}^2A_{m}^3\rangle$, to have a local realistic
model, for proofs see \cite{ww}, \cite{zb}. It is important to stress that the inequalities do not guarantee a local realistic model for the full set of probabilities describing the given experimental configuration.

\section{Bell inequalities involving only lower order correlations}

Our aim is to derive Bell inequalities which would reveal $N$-particle entanglement in the case when there are no $N$-qubit correlations. Thus, obviously such inequalities do not have to involve correlations of the highest order.
Still, our derivation will start with  specific initial $N$-particle inequalities, and in the next stage, by pruning from them $N$-particle correlations, we shall obtain the ones that we aim for.

The initial inequalities that we shall derive are saturated by half of the vertices of the Pitowsky polytope \cite{polytope,SLIWA}. For an introduction to this notion see Appendix.
We shall call them  {\em mirror inequalities}, as the other half saturate an analogous inequality with an opposite bound. Note that such  are the celebrated CHSH inequality \cite{CHSH}, and its generalizations to the multipartite cases, \cite{ww,wz,zb} (Werner-Wolf--Weinfurter-\.Zukowski-Brukner (WWW\.ZB) inequalities). A method to derive mirror Bell inequalities for $N$ qubits for  three observables per site was given in Ref. \cite{zukpoly}. Some explicit forms of such inequalities were obtained in Ref. \cite{wbz}. 

First we shall present our method for $N=3$.
\subsection{N=3}
As the first step we take the full algebraic expression used to derive the WWW\.ZB inequalities for $N=3$ given by (\ref{BELLEQ-4}). This reads
\begin{eqnarray}
\label{line1}
&\pm(A_1^1+A_2^1)(A_1^2+A_2^2)(A_1^\textbf{3}+A_2^3)&\nonumber\\
\label{line2}
&\pm(A_1^1+A_2^1)(A_1^2+A_2^2)(A_1^3-A_2^3)&\nonumber\\
\label{line3}
&\pm(A_1^1+A_2^1)(A_1^2-A_2^2)(A_1^3+A_2^3)&\nonumber\\
\label{line4}
&\pm(A_1^1+A_2^1)(A_1^2-A_2^2)(A_1^3-A_2^3)&\nonumber\\
\label{line5}
&\pm(A_1^1-A_2^1)(A_1^2+A_2^2)(A_1^3+A_2^3)&\nonumber\\
\label{line6}
&\pm(A_1^1-A_2^1)(A_1^2+A_2^2)(A_1^3-A_2^3)&\nonumber\\
\label{line7}
&\pm(A_1^1-A_2^1)(A_1^2-A_2^2)(A_1^3+A_2^3)&\nonumber\\
\label{line8}
&\pm(A_1^1-A_2^1)(A_1^2-A_2^2)(A_1^3-A_2^3)=\pm 8&
\end{eqnarray}

We now consider the second and the sixth line and observe that they contribute $\pm 8$ whenever $A_1^2=A_2^2$ and $A_1^3=-A_2^3$, regardless the relation between $A_1^1$ and $A_2^1$. They can be hence jointly replaced by $\pm 2(A_1^2+A_2^2)(A_1^3-A_2^3)$. Similarly, the third and the fourth line can be replaced by $\pm 2(A_1^1+A_2^1)(A_1^2-A_2^2)$ and the fifth and the seventh--by 
$\pm 2(A_1^1-A_2^1)(A_1^3-A_2^3)$. Thus,  we have

\begin{eqnarray}
\label{ref1}
& \frac{1}{8}(A_1^1+A_2^1)(A_1^2+A_2^2)(A_1^3+A_2^3)&\nonumber\\
&\pm \frac{1}{4}(A_1^1+A_2^1)(A_1^2-A_2^2)\pm \frac{1}{4}(A_1^2+A_2^2)(A_1^3-A_2^3)&\nonumber\\
&\pm \frac{1}{4}(A_1^1-A_2^1)(A_1^3+A_2^3)&\nonumber\\
&\pm \frac{1}{8}(A_1^1-A_2^1)(A_1^2-A_2^2)(A_1^3-A_2^3)=\pm1.& \label{INEQ3}
\end{eqnarray}
After the usual averaging of \eqref{INEQ3}, see the previous Section,  we get a set of $2^5$ inequalities (first found in ref. \cite{SLIWA}, using a computer algorithm)
\begin{eqnarray}
&\pm \langle\frac{1}{8}(A_1^1+A_2^1)(A_1^2+A_2^2)(A_1^3+A_2^3)&\nonumber\\
&\pm \frac{1}{4}(A_1^1+A_2^1)(A_1^2-A_2^2)\pm \frac{1}{4}(A_1^2+A_2^2)(A_1^3-A_2^3)\nonumber\\
&\pm \frac{1}{4}(A_1^1-A_2^1)(A_1^3+A_2^3)\nonumber\\
&\pm \frac{1}{8}(A_1^1-A_2^1)(A_1^2-A_2^2)(A_1^3-A_2^3)\rangle\leq 1,& \label{INEQ3a}
\end{eqnarray}
which, just as in the case of WWW\.ZB ones, can be wrapped up to just one
\begin{eqnarray}
& |\langle\frac{1}{8}(A_1^1+A_2^1)(A_1^2+A_2^2)(A_1^3+A_2^3)\rangle|&\nonumber\\
&+|\langle\frac{1}{4}(A_1^1+A_2^1)(A_1^2-A_2^2)\rangle| +|\langle\frac{1}{4}(A_1^2+A_2^2)(A_1^3-A_2^3)\rangle|\nonumber\\
&+|\langle\frac{1}{4}(A_1^1-A_2^1)(A_1^3+A_2^3)\rangle|\nonumber\\
& +|\langle\frac{1}{8}(A_1^1-A_2^1)(A_1^2-A_2^2)(A_1^3-A_2^3)\rangle|\leq 1,& \label{MODUL3}
\end{eqnarray}
Unfortunately, Ineq. (\ref{MODUL3}) is not violated by any state. Still by extending the above method to more qubits we get ``relevant" inequalities. 

 Once we have an inequality like the one based on (\ref{INEQ3}), we can drop the first and the last term to get
 \begin{eqnarray}
&-1 \leq\pm \frac{1}{4}\left\langle(A_1^1+A_2^1)(A_1^2-A_2^2)\pm \frac{1}{4}(A_1^2+A_2^2)(A_1^3-A_2^3)\right.&\nonumber\\
&\left.\pm \frac{1}{4}(A_1^1-A_2^1)(A_1^3+A_2^3)\right\rangle
\leq1.& \label{INEQ33}
\end{eqnarray}
and therefore a set of inequalities encapsulated in the following non-linear one:
\begin{eqnarray}
&|\langle\frac{1}{4}(A_1^1+A_2^1)(A_1^2-A_2^2)\rangle| +|\langle\frac{1}{4}(A_1^2+A_2^2)(A_1^3-A_2^3)\rangle|\nonumber\\
&+|\langle\frac{1}{4}(A_1^1-A_2^1)(A_1^3+A_2^3)\rangle| \leq 1.& \label{MODUL33}
\end{eqnarray}
They involve only two-particle correlations. Still, all pairs of observers enter the inequality on equal footing. Of course, one can generate more inequalities by swapping $+$'es and $-$'es for a party, etc..

\subsection{N=5}
In a similar fashion we will derive an inequality for 5 qubits. More examples can be found in Appendix. 

The five party WWW\.ZB expression reads
\begin{eqnarray}
&\sum_{s_1,s_2,s_3,s_4,s_5=\pm 1}\nonumber\\
&\big|(A_1^1+s_1A_2^1)(A_1^2+s_2A_2^2)(A_1^3+s_3A_2^3)\nonumber\\
&\times (A_1^4+s_4A_2^4)(A_1^5+s_5A_2^5)\big|
=32.&
\end{eqnarray}
Always only one  term gives a nonzero contribution. 
In the first step we choose two terms with completely opposite signs. These can be e.g.  $s_i=+$, $i=1,...,5$ and $s_i=-$ (it will be handy here to denote $\pm1$ as $\pm$, as what we are really interested in is whether we deal with a  sum or  a difference). We leave them as they are, they will enter the final expression unchanged. We are left with 30 terms to work with. We find two terms which differ in signs only for the fifth partner, and have the same sequence of signs for the other four, say $s_1=s_2=s_3=-s_4=+$. 
We may  replace both such  terms with  $2|\langle(A_1^1+A_2^1)(A_1^2+A_2^2)(A_1^3+A_2^3)(A_1^4-A_2^4)\rangle|$, and the value of the right hand side would stay put. This is because
\begin{eqnarray}
&|\sum_{s_5=\pm}(A_1^1+A_2^1)(A_1^2+A_2^2)&\nonumber\\ 
&\times(A_1^3+A_2^3)(A_1^4-A_2^4)(A_1^5+s_5A_2^5)|&\nonumber \\
&=2|(A_1^1+A_2^1)(A_1^2+A_2^2)(A_1^3+A_2^3)(A_1^4-A_2^4)|.&\nonumber 
\end{eqnarray}
Next we make a cyclic permutation of partners, and find four more pairs of terms for which we can do a similar replacement. We are left with 20 terms. We choose again two terms which differ in sign only for the last partner, however with a different sign sequence for the other four, say  $s_1=-s_2=s_3=-s_4=+$. The sum of the  terms gives $2|(A_1^1+A_2^1)(A_1^2-A_2^2)(A_1^3+A_2^3)(A_1^4-A_2^4)|$, and this is the replacement for them. Again we can easily find 4 pairs of terms in which the role of the partners is cyclically permuted, and we apply the same replacement rule. We are left with 10 terms. All of them have  three or four  $-$ signs. We can choose to pair two terms that differ by sign for the last partner, and have $-s_1=-s_2=-s_3=s_4=+$, and we pair them into $2(A_1^1-A_2^1)(A_1^2-A_2^2)(A_1^3-A_2^3)(A_1^4+A_2^4)|$. The  eight terms are a cyclic permutation of the two considered above, and can be paired in the same way. Note that e.g. we could have taken the two $-s_1=-s_2=+s_3=-s_4=+$ terms. Still all 10 terms could be paired. This would give us a different algebraic expression leading another inequality. However, one must stress that not all choices made during such a reduction process can lead to a successful pairing of all 30 terms.

  When we average over many realizations of the experiment, we get
\begin{eqnarray}
\label{INEQ5A}
&\frac{1}{32}\big[
|\langle\prod_{n=1}^{5}(A_1^n+A_2^n)\rangle|&\nonumber\\
&+2\sum_{CP\{+++-0\}} + 2\sum_{CP\{+-+-0\}} +2\sum_{CP\{---+0\}}&\nonumber\\
&+|\langle\prod_{n=1}^{5}(A_1^n-A_2^n)\rangle|\big]\leq 1, &\nonumber\\
\end{eqnarray}
where e.g. $\sum_{CP\{+++-0\}}$ denotes the sum of  $$ (|\langle(A_1^1+A_2^1)(A_1^2+A_2^2)(A_1^3+A_2^3)(A_1^4-A_2^4)\rangle|$$ and 
all four other such terms in which the role of the five partners is cyclically permuted.

Note that, if one is, e.g., interested in some three-particle correlations, one can further continue the reduction of the WWW\.ZB expression. One can group terms like  
\begin{equation} 2(|\langle(A_1^1+A_2^1)(A_1^2+A_2^2)(A_1^3+A_2^3)(A_1^4-A_2^4)\rangle| \label{TERM}\end{equation}

 and  $$ 2(|\langle(A_1^1+A_2^1)(A_1^2-A_2^2)(A_1^3+A_2^3)(A_1^4-A_2^4)\rangle|$$ to get  $$ 4|\langle(A_1^1+A_2^1)(A_1^3+A_2^3)(A_1^4-A_2^4)\rangle|.$$ Notice the exponential dependence of the factor, in front of the expression, on the number of omitted brackets. As inequality (\ref{INEQ5A}) has 15 terms of the types like (\ref{TERM}), in the considered example, it is not possible with this procedure of pairing to eliminate all four particle correlations.
\subsection{General Case}
It is now straight-forward to generalize the derivation presented above to any number of observers. The general WWW\.ZB algebraic expression leading to an inequality for $N$ qubits is
\begin{eqnarray}
\label{wwwzbgen}
&\frac{1}{2^N}\times\nonumber\\
&\sum_{s_1,s_2,...=\pm 1}S(s_1,s_2,...)\prod_{n=1}^{N}(A_1^n+s_nA_2^n)=\pm 1.\nonumber\\
\end{eqnarray} 
Again, $S(s_1,s_2,...)$ is a sign function dependent on signs $s_1,s_2,...$. 
We leave in peace the  term with only $+$ signs, and the term with only $-$ sign.
Next, we can choose a subset of all  terms from the sum in Eq. (\ref{wwwzbgen}) distinguished by specifying a certain combination of signs, for example $s_1=-$, and $s_2=+$, $s_3=+$,..., $s_{N-1}=+$ . This set of two terms can be replaced by a single term, in which the brackets for the last party  are replaced by factor 2. E.g. in the case given above  all terms with specified signs will be substituted by $\frac{1}{2^{N-1}}(A_1^1-A_2^1)\prod_{n=2}^{N-1}(A_1^n+A_2^n)$, etc. One can decide to eliminate more brackets in the first step. Thus we can e.g. choose all terms with  $s_1=-$, and $s_2=+$, $s_3=+$,..., $s_{N-k}=+$.
This would lead to $\frac{1}{2^{N-k}}(A_1^1-A_2^1)\prod_{n=2}^{N-k}(A_1^n+A_2^n)$.
This is continued till all terms in the original WWW\.ZB algebraic expression are taken into account in the procedure. The averaging of the resulting algebraic expression follows the usual pattern.
Note that if we want to have roles of all parties symmetric under permutation, such a procedure may halt at some moment, before the aim is reached. Still, in such a case we would have inequality in which at least some higher order correlations are eliminated. Note further, that as the last step we can remove the  term with only $+$ signs, and the term with only $-$ sign, to get an inequality involving only a certain set of lower order correlations.

\subsection{Violations}

We searched for violations  inequalities with the role  of the observes permutation invariant. The inequalities that gave us interesting results are
\begin{eqnarray}
\label{INEQ5B}
&\frac{1}{32}\big[
|\langle\prod_{n=1}^{5}(A_1^n+A_2^n)\rangle|+2\sum_{CP\{+++-0\}} + 2\sum_{CP\{-++-0\}}&\nonumber\\ 
&+2\sum_{CP\{-+--0\}}+|\langle\prod_{n=1}^{5}(A_1^n-A_2^n)\rangle|\big]\leq 1, &\nonumber\\
\end{eqnarray}
and its sister one with the first and the last term removed.
 We took 
as local observables  
$\hat{A}_1^{i}=\cos\phi_i \sigma_z+\sin\phi_i \sigma_x,$
and
$\hat{A}_2^{i}=\cos\phi_i \sigma_z-\sin\phi_i \sigma_x$
 with all $\phi_i= \phi$. The maximal eigenvalue of  five-qubit Bell operators related to Ineq. (\ref{INEQ5B}) was found to be 1.97435 with $\phi_i=\frac{\pi}4$. 
In studying inequalities for $N=7$ and $N=9$ given in the Appendix, we assumed $\phi_i\equiv \phi$ and the maximal observed violation ratios were 1.84331, 2.18414 and 1.79497, respectively. Thus even in this constrained scenario, inequalities can be strongly violated.

A five-qubit state giving the maximal violation of the wrapped
inequality (\ref{INEQ5B}), for Pauli observables constrained to the $XZ$ plane, however, without the $\phi_i=\phi$ symmetry constraint, reads
\begin{widetext}
\begin{eqnarray}
\label{PSI2}
&|\psi\rangle\approx a|00000\rangle
+b(|00011\rangle+|00110\rangle
+|01100\rangle+|11000\rangle+|10001\rangle)&\nonumber\\
&+c(|00101\rangle+|01010\rangle
+|10100\rangle+|01001\rangle+|10010\rangle)
+d(|00111\rangle+|01110\rangle
+|11100\rangle+|11001\rangle+|10011\rangle)&\nonumber\\
&+e(|01011\rangle+|10110\rangle
-|01101\rangle+|11010\rangle+|10101\rangle)
+f(|01111\rangle+|10111\rangle
+|11011\rangle+|11101\rangle+|11110\rangle)&\nonumber\\
&+g|11111\rangle,&
\end{eqnarray}
\end{widetext}
where $a=0.462854, b=0.161096, c=0.19409, d=0.181191, e=0.220891, f=0.107669$, and $g=0.039699$. 
The state is put here in the simplest form, a generalized Schmidt decomposition as defined in Ref. \cite{SUDBERY}. The state has been obtained by taking all possible pure five-qubit states and choosing for them the optimal observables to violate the Ineq. (\ref{INEQ5A}). Subsequently, local bases were chosen such that components with one $|1\rangle$ vanish. The procedure involved linear optimization procedure.

We can apply to this state a $NOT$ map, 
which flips the Bloch vector of a qubit state ($\sigma_a\stackrel{NOT}{\rightarrow}-\sigma_a, a=x,y,z$), to all qubits. Such a map is  composed of unitary rotations and a {\em full} transposition, and thus it is  positive one.  If we take an equal mixture $\rho=\frac{1}{2}(|\psi\rangle\langle\psi|+NOT^{\otimes 5}(|\psi\rangle\langle\psi|))$, the corresponding mean value reads 1.806, still much above 1. Notice that $\rho$ possesses the same 4-party correlations as $|\psi\rangle\langle\psi|$, but it is globally uncorrelated. Hence it does not violate {\em any}  WWW\.ZB inequality for five qubits. However, as shown in Ref. \cite{WIESNIAKMARU}  such a high violation of Ineq. (\ref{INEQ5B}) must follow from genuine five-partite entanglement. Also, note that this operation is equivalent to removing 5-qubit correlations from the inequality, and therefore moving to an inequality involving only four-partite correlations. Thus it can detect genuine five-particle entanglement, which has vanishing five-particle correlation functions.

\section{Closing remarks}
The presented method leads  to multipartite multisetting Bell inequalities which are an extension of the WWW\.ZB ones to multiorder correlations. The actual construction was inspired by Ref. \cite{wbz}. 
Note that characteristic trait of the new inequalities is that the lower-order terms as a whole take into account all observers. E.g., in ineq. \eqref{MODUL3} in each lower-order term one of the observers is excluded, however no observer is excluded from all such terms. In this way the new inequalities are sensitive to entanglement between all subsystems. The WWW\.ZB inequalities for $N-1$ observers cannot detect $N$ particle entanglement.

 Interesting open questions are, e.g., how many different permutation invariant inequalities can be constructed  for given $N$, or what other symmetries can be imposed? Recently 
a new class of multisetting Bell inequalities for three qubits was derived \cite{morenonlocality}. Establishing a relation of results of  \cite{morenonlocality} to the ones presented here could lead to interesting results. 
As the inequalities are violated by states that do not violate standard inequalities, this opens a possibility of a direct application of such states in quantum information protocols. Specific communication complexity reducing protocols, or some quantum games, are first that one can think of, but  more may lay in store.

The work is part of EU program QESSENCE
(Contract No.248095) and  MNiSW (NCN) Grant no. N202 208538.
MN is supported by the International PhD Project "Physics of future quantum-based information technologies": grant MPD/2009-3/4 of Foundation for Polish Science. MW was supported by a University of Gda\'nsk grant BW-538-5400-0624-1 at the early stage and by project HOMING PLUS from the Foundation for Polish Science at later stages. We are indebted to the Anonymous Referee who strongly suggested that we should present a simpler version of a derivation of the inequalities than the one of the initial manuscript, \cite{ARXIV}.

\appendix

\section{Further results}

\subsection{Examples of inequalities associated with 7 and 9 particles}

For seven qubits we have found the following  inequality

\begin{eqnarray}
&\frac{1}{128}\big[|\langle\prod_{n=1}^7(A_1^i+A_2^i)\rangle|\nonumber\\
&+2\sum_{CP(+++++-0)}+2\sum_{CP(++-++-0)}\nonumber\\
&+2\sum_{CP(+-+-++0)}+2\sum_{CP(-+--+-0)}\nonumber\\
&+2\sum_{CP(--+-++0)}+2\sum_{CP(+++---0)}\nonumber\\
&+2\sum_{CP(+-+---0)}+2\sum_{CP(-----+0)}\nonumber\\
&+|\langle\prod_{n=1}^7(A_1^i-A_2^i)\rangle|\big]\leq 1.\nonumber\\
\end{eqnarray}

For nine, which is a non-prime number

\begin{widetext}
\begin{eqnarray}
&\frac{1}{512}\big[|\langle\prod_{n=1}^9(A_1^i+A_2^i)\rangle|
+2\sum_{CP(+++++++-0)}+2\sum_{CP(-++++++-0)}+2\sum_{CP(-+++++-+0)}+2\sum_{CP(-++++-++0)}\nonumber\\
&+2\sum_{CP(-+++++--0)}+2\sum_{CP(-++++-+-0)}+2\sum_{CP(-++++--+0)}+2\sum_{CP(-++++---0)}\nonumber\\
&+2\sum_{CP(-+++-++-0)}+2\sum_{CP(-+++-+-+0)}+2\sum_{CP(-+++--+-0)}+2\sum_{CP(-+++-+--0)}\nonumber\\
&+2\sum_{CP(-+++--++0)}+2\sum_{CP(-+++----0)}+2\sum_{CP(-+++---+0)}+2\sum_{CP(-++-++--0)}\nonumber\\
&+2\sum_{CP(-++-+-+-0)}+2\sum_{CP(-++-+---0)}+2\sum_{CP(-++--++-0)}+2\sum_{CP(-++-----0)}\nonumber\\
&+2\sum_{CP(-++---+-0)}+2\sum_{CP(-++-+--+0)}+2\sum_{CP(--++--+-0)}+2\sum_{CP(-+-+--+-0)}\nonumber\\
&+2\sum_{CP(-+-+----0)}+2\sum_{CP(--+--+-+0)}+2\sum_{CP(----+--+0)}+2\sum_{CP(---+----0)}\nonumber\\
&+|\langle(A_1^1+A_2^1)(A_1^2+A_2^2)(A_1^3-A_2^3)(A_1^4+A_2^4)(A_1^5+A_2^5)(A_1^6-A_2^6)(A_1^7+A_2^7)(A_1^8+A_2^8)(A_1^9-A_2^9)\rangle|\nonumber\\
&+|\langle(A_1^1+A_2^1)(A_1^2-A_2^2)(A_1^3+A_2^3)(A_1^4+A_2^4)(A_1^5-A_2^5)(A_1^6+A_2^6)(A_1^7+A_2^7)(A_1^8-A_2^8)(A_1^9+A_2^9)\rangle|\nonumber\\
&+|\langle(A_1^1-A_2^1)(A_1^2+A_2^2)(A_1^3+A_2^3)(A_1^4-A_2^4)(A_1^5+A_2^5)(A_1^6+A_2^6)(A_1^7-A_2^7)(A_1^8+A_2^8)(A_1^9+A_2^9)\rangle|\nonumber\\
&+|\langle(A_1^1-A_2^1)(A_1^2-A_2^2)(A_1^3+A_2^3)(A_1^4-A_2^4)(A_1^5-A_2^5)(A_1^6+A_2^6)(A_1^7-A_2^7)(A_1^8-A_2^8)(A_1^9+A_2^9)\rangle|\nonumber\\
&+|\langle(A_1^1+A_2^1)(A_1^2+A_2^2)(A_1^3-A_2^3)(A_1^4+A_2^4)(A_1^5+A_2^5)(A_1^6-A_2^6)(A_1^7+A_2^7)(A_1^8+A_2^8)(A_1^9-A_2^9)\rangle|\nonumber\\
&+|\langle(A_1^1+A_2^1)(A_1^2+A_2^2)(A_1^3-A_2^3)(A_1^4+A_2^4)(A_1^5+A_2^5)(A_1^6-A_2^6)(A_1^7+A_2^7)(A_1^8+A_2^8)(A_1^9-A_2^9)\rangle|\nonumber\\
&+|\langle\prod_{n=1}^9(A_1^i-A_2^i)\rangle|\big]\leq 1\\
\end{eqnarray}
and
\begin{eqnarray}
\label{Athree}
&\frac{1}{512}\big[|\langle\prod_{n=1}^9(A_1^i+A_2^i)\rangle|+8\sum_{CP(+++++-000)}+8\sum_{CP(-++++-000)}+8\sum_{CP(-+++--000)}\nonumber\\
&+8\sum_{CP(-++---000)}
+8\sum_{CP(-++0-+-00)}+8\sum_{CP(-++-+-000)}+8\sum_{CP(--+-0-0-0)}\nonumber\\
&+|\langle(A_1^1+A_2^1)(A_1^2+A_2^2)(A_1^3-A_2^3)(A_1^4+A_2^4)(A_1^5+A_2^5)(A_1^6-A_2^6)(A_1^7+A_2^7)(A_1^8+A_2^8)(A_1^9-A_2^9)\rangle|\nonumber\\
&+|\langle(A_1^1+A_2^1)(A_1^2-A_2^2)(A_1^3+A_2^3)(A_1^4+A_2^4)(A_1^5-A_2^5)(A_1^6+A_2^6)(A_1^7+A_2^7)(A_1^8-A_2^8)(A_1^9+A_2^9)\rangle|\nonumber\\
&+|\langle(A_1^1-A_2^1)(A_1^2+A_2^2)(A_1^3+A_2^3)(A_1^4-A_2^4)(A_1^5+A_2^5)(A_1^6+A_2^6)(A_1^7-A_2^7)(A_1^8+A_2^8)(A_1^9+A_2^9)\rangle|\nonumber\\
&+|\langle(A_1^1-A_2^1)(A_1^2-A_2^2)(A_1^3+A_2^3)(A_1^4-A_2^4)(A_1^5-A_2^5)(A_1^6+A_2^6)(A_1^7-A_2^7)(A_1^8-A_2^8)(A_1^9+A_2^9)\rangle|\nonumber\\
&+|\langle(A_1^1+A_2^1)(A_1^2+A_2^2)(A_1^3-A_2^3)(A_1^4+A_2^4)(A_1^5+A_2^5)(A_1^6-A_2^6)(A_1^7+A_2^7)(A_1^8+A_2^8)(A_1^9-A_2^9)\rangle|\nonumber\\
&+|\langle(A_1^1+A_2^1)(A_1^2+A_2^2)(A_1^3-A_2^3)(A_1^4+A_2^4)(A_1^5+A_2^5)(A_1^6-A_2^6)(A_1^7+A_2^7)(A_1^8+A_2^8)(A_1^9-A_2^9)\rangle|&\nonumber\\
&+|\langle\prod_{n=1}^9(A_1^i+A_2^i)\rangle|\big]\leq 1\nonumber\\
\end{eqnarray}

\end{widetext}
The sets defining the allowed permutations, were found with a computer algorithm following the patter  described in the main text. We have started with a seed containing two $N$-symbol strings, namely the ones with only $+$'es and $-$'es. Then, a sequence of $+$'es, $-$'es, and $0$'s was drawn at random and tested for desired properties: if the number of zeros, $k$, equal to 3 for the problem leading to (\ref{Athree}), and 1 otherwise. We also demand that the a candidate sequence has at least one $+/-$ mismatch with all vectors in the temporary set. When these criteria are met, all the cyclic permutations of the candidate are added to the temporary set. These steps were repeated until the set was complete, which was indicated by its cardinality. Since the algorithm is stochastic, various, non-equivalent sets can be generated.
  
\subsection{Derivation of sufficient conditions to satisfy the inequalities}
Let us now briefly present a method to establish  a necessary condition to violate the inequalities. The conditions are very easy to reach, mainly due to the fact that the inequalities have the trait that they can be put in the form of a single nonlinear inequality.  We shall use the approach of Ref. \cite{zb}. The arbitrary mixed state for N qubits can be written as

\begin{eqnarray}
\rho = \frac{1}{2^N} \sum_{k_1,\dots,k_N = 0}^3{T_{k_1,\dots,k_N}\sigma_{k_1}^1\otimes \dots \otimes \sigma_{k_N}^N},
\end{eqnarray}
where $\sigma_0^i$ is the identity operator and $\sigma_{k_i}^i$ are the Pauli operators for the three orthogonal directions $k_i=1,2,3$. The set of real coefficients $T_{k_1,\dots,k_N}$, with $k_i=0,1,2,3$ form the so-called correlation tensor $T$. 
In the course of constructing the inequalities the pairs of local observables occur either as $\frac{1}{2}(\hat{A}^i_1+\hat{A}^i_2)$ or $\frac{1}{2}(\hat{A}^i_1-\hat{A}^i_2)$. 
Take any possible pair of observables for the party $i$, that is $\hat{A}_1^i=\vec{n}_1^i\cdot\vec{\sigma}$ and $\hat{A}_2^i=\vec{n}_2^i\cdot\vec{\sigma}$, where $\vec{n}_k^i$ are unit vectors, and $\vec{\sigma}$ is a quasi-vector formed of Pauli operators for $i$.  As  $\vec{n}^i_1+\vec{n}^i_2$ and $\vec{n}^i_1-\vec{n}^i_2$ form an orthogonal pair, after some algebra we see that we can put
 $\frac{1}{2}(\hat{A}^i_1+\hat{A}^i_2)=\cos\theta_i \vec{w}^i\cdot\vec{\sigma}$ and $\frac{1}{2}(\hat{A}^i_1-\hat{A}^i_2)=\sin\theta_i \vec{w}^{\perp i}\cdot\vec{\sigma}$, where $\phi_i$ is certain angle, dependent on $\vec{n}_1^i$ and $\vec{n}_2^i$, and $\vec{w}^i$ and $\vec{w}^{\perp i}$ form an orthonormal pair. For simplicity we put $\vec{w}^i=\vec{x}^i$ and $\vec{w}^{\perp i} =\vec{z}^i$. Then we have $\frac{1}{2}(\hat{A}^i_1+\hat{A}^i_2)=\cos\theta_i \: \sigma_1^i$ and $\frac{1}{2}(\hat{A}^i_1-\hat{A}^i_2)=\sin\theta_i  \:\sigma_3^i$.
We now can rewrite our Bell expression for the wrapping inequality with moduli, such as \eqref{MODUL3},  as follows
 \begin{eqnarray}
& |\text{Tr}[\rho \; (\cos\theta_1 \: \sigma_1^1\;\cos\theta_2 \: \sigma_1^2\;\cos\theta_3 \: \sigma_1^3)]|&\nonumber\\
&+|\text{Tr}[\rho \; (\cos\theta_1 \: \sigma_1^1\;\sin\theta_2  \:\sigma_3^2\;\sigma_0^3)]|\nonumber\\ 
&+|\text{Tr}[\rho \; (\sigma_0^1\;\cos\theta_2 \: \sigma_1^2\;\sin\theta_3  \:\sigma_3^3)]|\nonumber\\
&+|\text{Tr}[\rho \; (\sin\theta_1  \:\sigma_3^1\;\sigma_0^2\;\cos\theta_3 \: \sigma_1^3)]|\nonumber\\
& +|\text{Tr}[\rho \; (\sin\theta_1  \:\sigma_3^1\;\sin\theta_2  \:\sigma_3^2\;\sin\theta_3  \:\sigma_3^3)]|\leq 1,& \label{MODUL4}
\end{eqnarray}
After taking the trace equation \eqref{MODUL4} becomes
\begin{eqnarray}
& |\cos\theta_1 \: \cos\theta_2 \: \cos\theta_3 \: T_{111}|
+|\cos\theta_1 \: \sin\theta_2 \: T_{130}|&\nonumber\\
&+|\cos\theta_2 \: \sin\theta_3 \: T_{013}|
+|\sin\theta_1 \: \cos\theta_3 \: T_{301}|\nonumber\\
&+|\sin\theta_1 \: \sin\theta_2 \: \sin\theta_3 \: T_{333}|\leq 1. & \label{MODUL5}
\end{eqnarray}

For a reason, which will be clear after few lines, we will write \eqref{MODUL5} as a scalar product between two vectors, one containing only trigonometric functions, and the other moduli of elements of the correlation tensor, which could be written in general as $T_{k_1,\dots,k_N}=\text{Tr}[\rho\;(\sigma_{k_1}\otimes\ \dots \otimes \sigma_{k_N})]$. Then we have the following inequality
\begin{eqnarray}
&(|\cos\theta_1\cos\theta_2\cos\theta_3|,|\cos\theta_1\sin\theta_2|,&\nonumber\\
&|\cos\theta_2\sin\theta_3|,|\sin\theta_1\cos\theta_3|,|\sin\theta_1\sin\theta_2\sin\theta_3|)&\nonumber\\
&\cdot(|T_{111}|,|T_{130}|,|T_{013}|,|T_{301}|,|T_{333}|)\leq 1.&
\end{eqnarray} 
It is now easy to see that the norm of the first vector is at most 1. 
Thus,  Cauchy-Schwartz inequality gives
\begin{equation}
T_{111}^2+T_{130}^2+T_{013}^2+T_{301}^2+T_{333}^2\leq 1,
\end{equation}
as a sufficient condition to satisfy the inequality.
If it holds for a given three-qubit state for all local Cartesian bases, the nonlinear inequality (10) cannot be violated by this state (unfortunately this is so for {\em any} state). 

Similarly, the sufficient condition for a state to satisfy the inequality 
(\ref{INEQ5B})
 is given by
\begin{eqnarray}
\label{condi}
&\big( T_{33333}^2+T_{11111}^2\big)+T_{33310}^2+T_{03331}^2+T_{10333}^2+T_{31033}^2&\nonumber\\
+&T_{33103}^2+T_{13310}^2+T_{01331}^2+T_{10133}^2+T_{31013}^2+T_{33101}^2&\nonumber\\
+&T_{13110}^2+T_{01311}^2+T_{10131}^2+T_{11013}^2+T_{31101}^2\leq1.
\end{eqnarray}
In the case of the derivative inequality, which does not involve five particle correlations, one simply drops all terms without index $0$, that is the first two ones. This why they are in a  bracket. 

\subsection{Pitowsky polytope}

A geometric approach to the problem of Bell inequalities based on a well established fact that the set of all possible  probabilities of experimental results, for experiments involving a finite number of settings, describable by local realistic theories forms a convex ``Pitowsky'' polytope, \cite{polytope,SLIWA}.  Take probabilities for the three particle case described above. The assumption of existence of local realistic models allows one to introduce an underlying 
probability distribution, describing the behavior of the $N$ particles in all possible situations. That is we assume that there exists $p_{LR}(A_{1}^{1},A_{2}^{1}, A_{1}^{2},A_{2}^{2}, A_{1}^{3},A_{2}^{3})$, and it it a proper probability distribution having all usual features (non-negativity, and summation over all possible values giving $1$). The Pitowsky polytope in such a case is a geometric construction which formed by observation that such a probability distribution can always be treated as a convex combination of trivial deterministic cases in which only one configuration of values of $(A_{1}^{1},A_{2}^{1}, A_{1}^{2},A_{2}^{2}, A_{1}^{3},A_{2}^{3})$ is occurring (that is has probability $1$) and all other ones have probability $0$. Such a trivial deterministic probability forms a vertex of the Pitowsky  polytope (in our example there are $64$ such vertices). As we can map one-to-one the deterministic cases with the configurations for which they give probability $1$, we can form a polytope in a $26$ dimensional vector space. 
In the original approach one used yes/no representation of the results, that is instead of $A_{m}^{k}$ the values $r_{m}^{k}=\frac{1}{2}(1- A_{m}^{k})$, equal either $0$ or $1$, and the $26$ dimensional vector was ascribed components, which were in the form of a sequence formed out of all specific values of $r_{m}^{k}$, six of them, values of all two observer products of these for  pairs for different observers $A_{n}^{k}A_{m}^{k'}$, for $k\neq k'$, twelve of them, and finally all product of the kind $r_{l}^{1}r_{n}^{2}r_{m}^{3}$, eight of them. 
 This is the geometric picture of the full set of all possible local realistic models for the given experimental configurations (here three observers, two observables for each observer, dichotomic results).  
 
 An  equivalent approach, which we think is technically more useful, can be formulated as follows. It allows one to use some powerful tools of algebraic geometry, especially the properties of tensor products of vector spaces. In the case presented above we would link the deterministic probability vertex with a 27 dimensional vector, or factorisable tensor of the form  
\begin{equation}
\hat{h}(\vec{A})=(A^1_1,A^1_2,1)\otimes(A^2_1,A^2_2,1)\otimes(A^3_1,A^3_2,1),\label{REAL3}
\end{equation} 
		where $\vec{A}=(A^1_1,A^1_2,A^2_1,A^2_2,A^3_1,A^3_2).$
They are in a one-to-one way relation with the vertices of the Pitowsky  polytope of the problem. As one of the components of the tensor product is always equal to  $1$, the vertices span a $26$ dimensional hyperplane which contains a Pitowsky polytope (of a new, but equivalent, kind).  The polytope is of course the full set of all possible {\em convex} combinations of the tensor-vertices. Thus we shall call them simply {\em vertices}. The tensor component  always equal to $1$ plays only a technical role. The tensor product structure of the vertices simplifies the analysis, and proved to be very useful, in our work on the new inequalities (see the initial version of the work in Arxiv.org \cite{ARXIV}).

\newpage

\newpage

\end{document}